# Lessons from commissioning of the cryogenic system for the Short-Baseline Neutrino Detector at Fermilab

**Frederick Schwartz[a], Roberto Acciarri[a], Johan Bremer[b], Roza Doubnik[a], Caroline Fabre[b], Michael Geynisman[a], Claudio Montanari[a,c], Monica Nunes[a], Trevor Nichols[a], William Scofield[a], Zach West[a], Peter Wilson[a]**

[a] Fermi National Accelerator Laboratory
Batavia, IL 60510 United States, hope@fnal.gov
[b] CERN, 1211 Geneva-23, Switzerland
[c] Istituto Nazionale di Fisica Nucleare (INFN), Pavia, Italy

## Abstract

Results from commissioning and first year of operations of the cryogenic system of the Short-Baseline Neutrino Detector (SBND) and its membrane cryostat installed at the Fermi National Accelerator Laboratory are described. The SBND detector is installed in a 200 m$^3$ membrane cryostat filled with liquid argon, which serves both as target and as active media. For the correct operation of the detector, the liquid argon must be kept in very stable thermal conditions while the contamination of electronegative impurities must be consistently kept at the level of small fractions of parts per billion. The detector is operated in Booster Neutrino Beams (BNB) at Fermilab for the search of sterile neutrinos and measurements of neutrino-argon cross sections. The cryostat and the cryogenic systems also serve as prototypes for the much larger equipment to be used for the LBNF/DUNE experiment. Since its installation in 2018-2023 and cooldown in spring of 2024, the cryostat and the cryogenic system have been commissioned to support the detector operations. The lessons learned through installation, testing, commissioning, cooldown, and initial operations are described.

## 1. Introduction

The Short-Baseline Neutrino (SBN) program in the Fermilab Booster Neutrino Beam (BNB) searches for short-baseline neutrino oscillations and the existence of light sterile neutrinos (Acciarri R. et al., 2015). Presently, it relies on the deployment and operations of two high precision neutrino detectors at different distances along a single high-intensity neutrino beam, each using Liquid Argon (LAr) Time Projecting Chamber (TPC) technology. Both LAr-TPC detectors, namely SBN's Far Detector (SBN-FD, also known as ICARUS-T600 or Icarus for short) and SBN's Near Detector (SBND) are essentially cryostats filled with high purity liquid argon and equipped with electronics that measure the ionization charge and scintillation light produced by the interaction between argon atoms and neutrinos. The Icarus detector has been operational since Feb 2020 (Geynisman et al., 2021, 2023). The SBND detector and its cryogenic system (Geynisman et al., 2017) was commissioned in 2024. While the SBND cryostat is built with the same membrane technology as two existing ProtoDUNE cryostats at the European Organization for Nuclear Research (CERN), it is the only one now commissioned and operated for collecting scientific data. The cryostat and the cryogenic system of the SBND at Fermilab were designed, built, and installed by a scientific collaboration between CERN and Fermilab. This paper briefly describes the installation, commissioning, and transition to operations of the SBND cryostat and the cryogenic and purification system as related to two ultimate design requirements: a) stability of cryogenic system as related to maintaining operations of the SBND TPC and other scientific equipment and b) reaching and maintaining the purity level of the liquid argon in the detector volume above 3 milliseconds electron lifetime.

## 2. Description of the SBND cryostat and cryogenic system

The SBND cryogenic system was designed to support operations of the membrane cryostat with its 112-ton active mass liquid argon TPC with ionization charge and scintillation light detection for the reconstruction of neutrino interactions. This membrane cryostat provided by CERN and Fermilab (Figure 1) utilizes an external warm structure with open top to support an 800-mm thick cold package comprised of an inner metallic membrane, secondary containment, and insulation. The top, from which the detector is suspended, is sealed by four plates, which provide the necessary scientific and cryogenic penetrations through an 800-mm insulation space. The warm structure, designed and installed by CERN with Fermilab assistance, consists of large vertical and horizontal beams and is singularly responsible for supporting all internal and external loads. Internal forces are transferred through the flexible 1.2 mm thick welded stainless steel membrane and load bearing insulation to this warm steel structure. The cold package, designed by Gas Transport Technology GTT, France and installed by CERNs subcontractor Gabadi, Spain, ensures leak tightness and necessary thermal flexibility provided by a network of orthogonal corrugations in the primary membrane. The inner dimensions of the cryostat are 5.42 m high, 7.02 m length, and 5.20 m wide, resulting in a total volume of ~200 m$^3$. In the event of a leak through the primary membrane, the secondary membrane and the outer structure ensure additionally the gas and liquid tightness. The secondary barrier is fabricated from a composite material and is incorporated into the insulation panels. The flexible tape maintains the integrity of the secondary barrier by bonding of strips between insulating panels. Thermal insulation is provided by the foam insulation in between the membrane and the outer structure and is designed to limit the thermal exchanges through the sides of the tank to 10 W/m². This insulation space is maintained under positive nitrogen atmosphere of 4-10 mbarg to avoid entrance of oxygen and water into this volume.

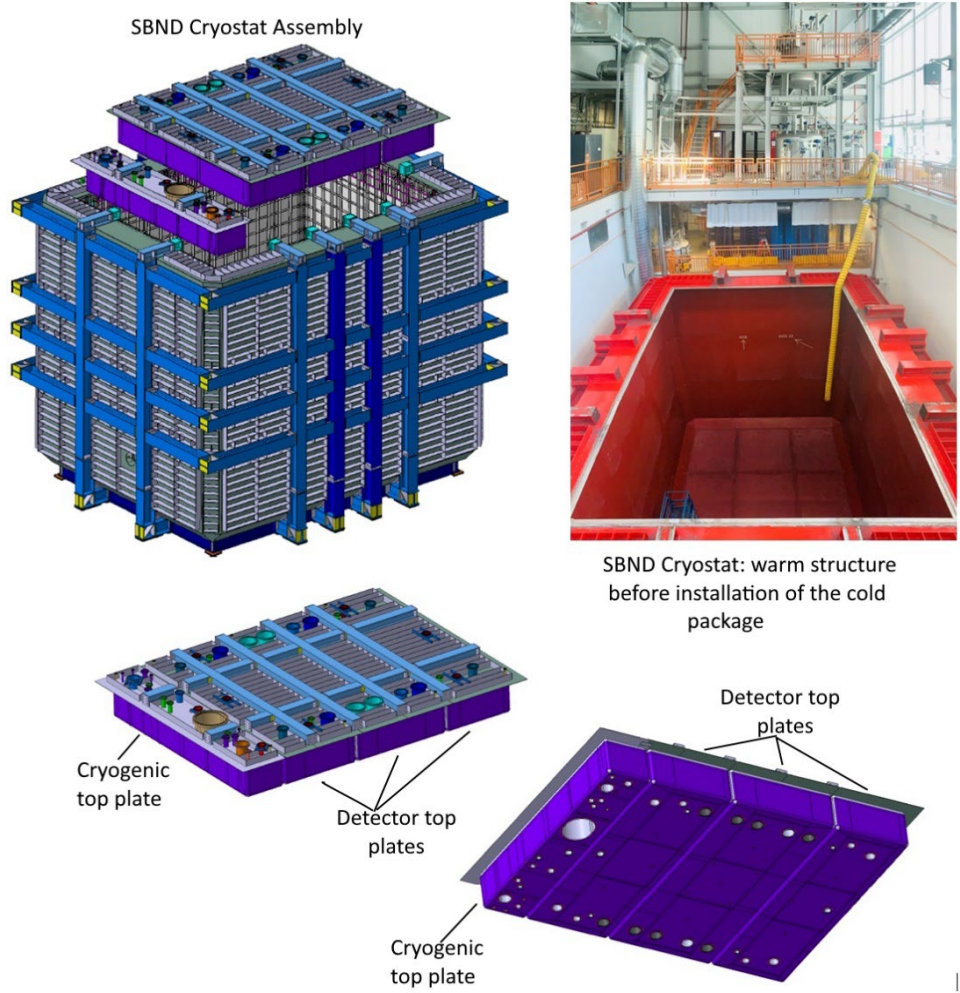

**Figure 1: SBND Cryostat**

The top plate assembly, which completes the structural and pressure boundaries of the cryostat, consists of a cryogenic section and a detector section, welded together. The cryogenic section has 20 penetrations interconnecting the cryogenic system and the cryostat, including liquid and gas argon transfer lines, gas venting, level and pressure sensors, and a manhole. The detector section contains all scientific penetrations terminated with conflat flanges incorporating feedthroughs for passing the detector wiring from the cryostat volume to the outside. Impurities outgassed from the composite materials and wiring inside the scientific penetrations are removed by flowing argon gas from the ullage of the cryostat to a condenser through the warm filtration system.

The cryogenic system (Figures 2, 3) includes 36 m$^3$ liquid nitrogen and 30 m$^3$ liquid argon storage dewars, 14 kW 1.1 m$^3$ argon/nitrogen condenser, a 0.8 m$^3$ argon phase separator, three 0.5 m$^3$ liquid argon and one 0.06 m$^3$ warm argon gas filters, all filled with activated copper (Cu-0226) and 4A molecular sieve, two liquid argon circulation and one condenser recirculation pumps, regeneration, gas analysis and purity measurement equipment, as well as cryogenic piping. The condenser, phase separator, cold and warm filters and Barber-Nichols argon circulation pumps, are packaged in vacuum insulated valve boxes designed, manufactured and installed by Demaco, Holland per contract with CERN. All external cryogenic equipment, including the dewars, regeneration, gas analysis, gas handling and venting, as well as safety and control systems, were designed and installed by Fermilab.

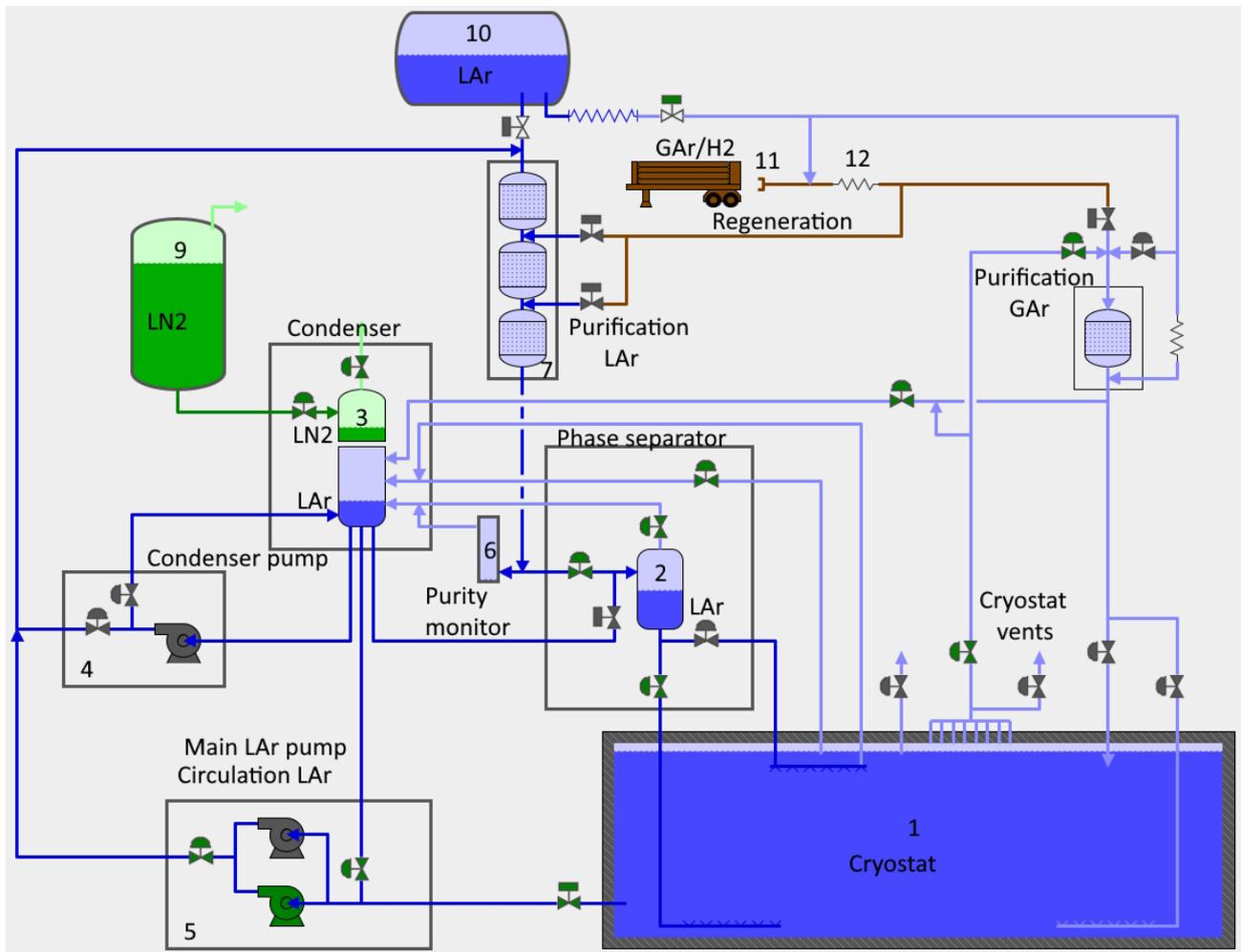

**Figure 2: SBND Cryogenic System Schematic**

1 – cryostat, 2 – argon phase separator, 3 – condenser, 4 - condenser pump, 5 – main circulation pumps, 6 – in-line purity monitor, 7 – liquid argon filter unit, 8 – gas argon filter unit, 9 – LN2 dewar, 10 – LAr dewar, 11 – Ar/H$_2$ trailer, 12 – regeneration station (not shown: 13 - gas analysis equipment and PLC-based safety and controls systems)

The cryogenic system was designed to support stable operations in all modes of operation, including purge, cooldown and fill of the cryostat, circulation of the liquid and gaseous argon via filters, phase separating of liquid argon before delivering it to the cryostat via transfer lines and distribution manifolds, reliquefying boiloff argon in the Ar/N$_2$ condenser to maintain stable pressure in the cryostat, regeneration of the copper and molecular sieve in the filters, purging of cryostat insulation spaces with gas nitrogen, and sampling of impurities in the argon throughout the system.

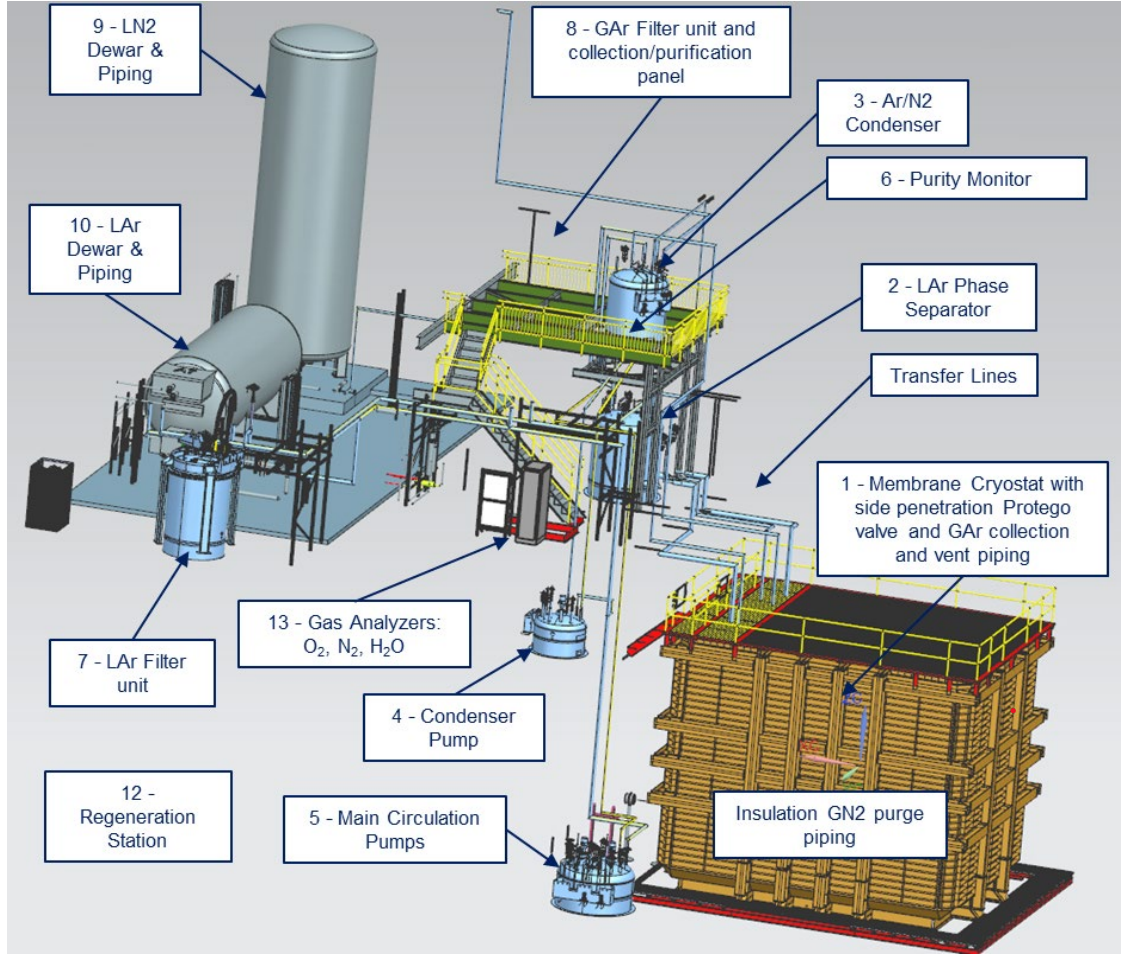

**Figure 3: SBND Cryogenic System Physical Layout**

The cryogenic system was designed to demonstrate a bulk electronegative impurity concentration ρ in argon being below 100 ppt (10$^{-10}$ parts) to achieve a free electron lifetime in argon τ=3 ms or better for the design drift time T=1.28 ms and attenuation A=34.7% (Equation 1, 2).

$$A\ [\%] = 100\ (1 - e^{-\frac{T}{\tau}})  \qquad\qquad Equation\ (1)$$

$$\tau\ [ms] \approx \frac{300\ [ms \cdot ppt\ Oxygen\ Equivalent]}{\rho\ [ppt\ Oxygen\ Equivalent]}  \qquad\qquad Equation\ (2)$$

As the SBND cryostat and its TPC serve to prototype for future LBNF/DUNE far detector installations (B. Abi et al., 2020), which will need a 2.5 ms design drift time, it is important to demonstrate a free electron lifetime above 6 ms for comparable attenuation of 34%. If a lifetime is shown to be 10 ms, the attenuation drops to 22%.

### 3. Commissioning of the cryostat and the cryogenic system

The commissioning of the cryostat and the cryogenic system was performed in several stages after the completion of installation in late 2023 and a cryostat pressure test in October 2023. The initial drying of the cryostat was done with air at -40C dewpoint in January 2024. The cryogenic system was purified below 1 ppm by multicycle pumping to vacuum and pressurizing with argon. The liquid argon and warm filters were regenerated and verified to produce filtration for $O_2$ and $H_2O$ to low ppb range. Then, from January 24 through February 2 of 2024, the cryostat was "piston purged" with purified and heated argon by flowing it from the bottom of the cryostat and venting it to atmosphere at ~1.6 g/s. When the impurities in the vented argon dropped well below 1 ppm, every penetration to the cryostat and every bayonet in the transfer lines was vented to clean trapped volumes and then resealed or welded. The cooldown of the cryostat and TPC started on February 6, 2024, and was achieved through a combination of spraying argon from the atomizing nozzles on top distribution manifold and delivering argon to the liquid distribution manifold on the bottom of the cryostat. The objective was to maintain the temperature gradients in the TPC structure within < 10 K/m (vertically) or 50 K overall at a cooldown speed below 40 K/hr. The cooldown process continued for 3 days until the temperatures stabilized across the TPC structure (Figure 4). The fill of the cryostat with liquid argon was started immediately after cooldown and proceeded with delivery of 18 truckloads, each ~ 14000 litres, while transferring argon to the cryostat via the liquid argon filter and the argon phase separator while verifying the impurities in argon entering the cryostat to be in ppb range. The maximum temperature gradient never exceeded 40K over the full 4m height of the TPC.

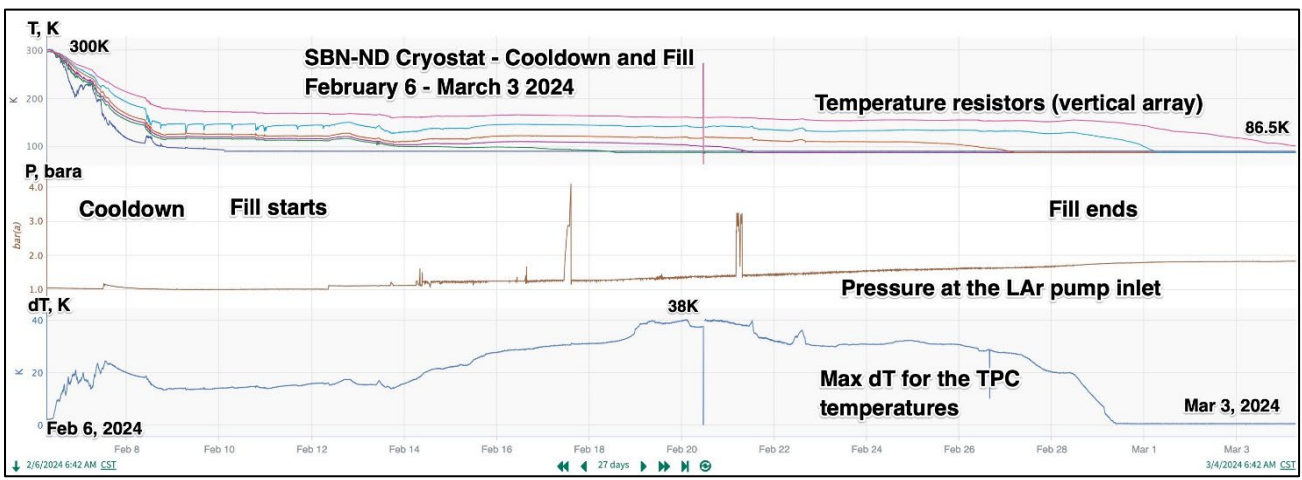

**Figure 4: Cooldown and Fill of the SBND Cryostat (Feb 6 – Mar 3, 2024)**

The condenser was commissioned on February 26, 2024, allowing the gaseous argon boiloff to be continuously collected, recondensed, and returned to the bottom of the cryostat rather than vented, greatly reducing the loss of argon for the filling process. The recirculation of the liquid argon through the filtration system and return to the cryostat at ~3.3 $m^3$/hr rate started on March 12. Many of the lessons learned from commissioning of the ICARUS cryogenic system at Fermilab (Geynisman M. et al., 2023), most importantly the need to vent argon to clear vapor locks in the "goosenecked" piping inside the valve boxes, were successfully implemented for the SBND cryogenic system.

The scientific subsystems of the TPC detector were in a commissioning stage starting with the completion of the cryostat fill in March 2024. The only significant upset for the TPC commissioning occurred while ramping the high voltage (HV) to its nominal -100 kV. During the investigation of these HV problems, an intervention under one of the electronics feedthrough flanges became necessary. The flange was safely opened on June 20, 2024, while the argon gas pressure was released and kept at atmospheric level to allow personnel, fitted with air masks, to reach into cold space of the cryostat and work below the flange. The cryogenic intervention itself was a unique experience and a success, but it didn't have the expected results for the detector at that moment. However, one week later while attempting to ramp up the HV through -35kV, the spikes and instabilities on the circuit did not reappear. The detector was then successfully ramped to nominal -100 kV on the cathode and has since been remaining stable, allowing for the transition to scientific data collection.

The bulk argon purity was measured with the purity monitor installed inside the cryostat. The purity monitors are not absolutely calibrated, which is why there is no absolute scale. However, the purity monitor indicated an increase of greater than 800% in free electron lifetime during initial cleanup of liquid argon being circulated and filtered in the liquid argon filter in March 2024 (Figure 5). The absolute value of the free electron lifetime was later measured with the TPC using cosmic tracks. This measurement is more sensitive than the purity monitors. This initial measurement was taken at a single point in time, July 24, 2024. The analysis shows that the purity was already better than 10 ms at that time, which far exceeds the design requirement of 3 ms. More measurements are expected to be taken by the time of publication.

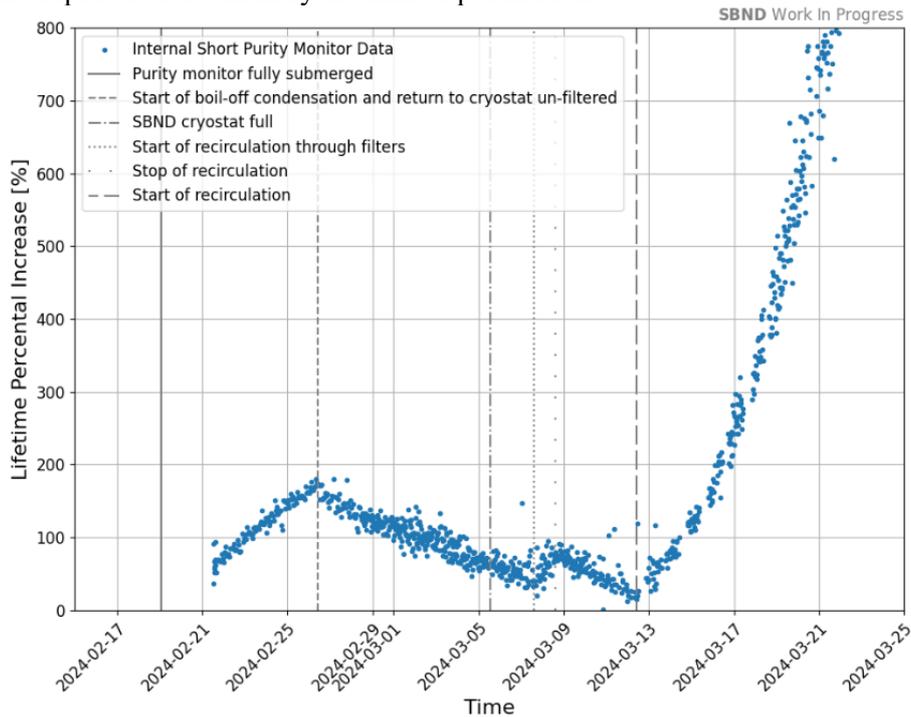

**Figure 5: Electron lifetime measured during initial argon cleanup in the SBND cryostat**

## 4. Conclusions

The SBND cryogenic and purification systems are now fully operational to support the scientific mission of the LArTPC detector in the Booster Neutrino Beam at Fermilab. Many useful lessons have been learned and improvements have been made to achieve this success, demonstrating the viability of membrane cryostat technology and design of cryogenic system for future LBNF/DUNE project. More details and physics results will be described in later publications.

## Acknowledgements

This manuscript has been authored by Fermi Forward Discovery Group, LLC under Contract No. 89243024CSC000002 with the U.S. Department of Energy, Office of Science, Office of High Energy Physics.